# Unveiling the curtain of superposition: Recent gedanken and laboratory experiments

E Cohen[1,2] and A C Elitzur[2]

[1] H.H. Wills Physics Laboratory, University of Bristol, Tyndall Avenue, Bristol, BS8 1TL, U.K
[2] Iyar, The Israeli Institute for Advanced Research, POB 651, Zichron Ya'akov 3095303, Israel

Email: eliahu.cohen@bristol.ac.uk

**Abstract.** What is the true meaning of quantum superposition? Can a particle genuinely reside in several places simultaneously? These questions lie at the heart of this paper which presents an updated survey of some important stages in the evolution of the three-boxes paradox, as well as novel conclusions drawn from it. We begin with the original thought experiment of Aharonov and Vaidman, and proceed to its non-counterfactual version. The latter was recently realized by Okamoto and Takeuchi using a quantum router. We then outline a dynamic version of this experiment, where a particle is shown to "disappear" and "re-appear" during the time evolution of the system. This surprising prediction based on self-cancellation of weak values is directly related to our notion of Quantum Oblivion. Finally, we present the non-counterfactual version of this disappearing-reappearing experiment. Within the near future, this last version of the experiment is likely to be realized in the lab, proving the existence of exotic hitherto unknown forms of superposition. With the aid of Bell's theorem, we prove the inherent nonlocality and nontemporality underlying such pre- and post-selected systems, rendering anomalous weak values ontologically real.

## 1. Preface

(He looked under the bed, in the chimney, in the cupboard; – nobody. He could not understand how he got in, or how he escaped.
Hoffmann. – Nocturnal Tales)

Oh! how often have I heard and seen him, Scarbo, when at midnight the moon glitters in the sky like a silver shield on an azure banner strewn with golden bees.

How often have I heard his laughter buzz in the shadow of my alcove, and his fingernail grate on the silk of the curtains of my bed!

How often have I seen him alight on the floor, pirouette on a foot and roll through the room like the spindle fallen from the wand of a sorceress!

Do I think him vanished then? the dwarf grows between the moon and me like the belfry of a gothic cathedral, a golden bell shakes on his pointed cap!

But soon his body becomes blue, translucent like the wax of a candle, his face pales like the wax of a candle end – and suddenly he is extinguished.

*Scarbo (translated to English), from Gaspard de la nuit / Aloysius Bertrand*

## 2. Introduction

In a famous letter Einstein has replied to a bewildered letter from Louis de Broglie's PhD advisor. Paul Langevin did not know what to think about the youngster's idea which has generalized Einstein's earlier Nobel-winning discovery that light, hitherto regarded as a wave, was in fact composed of particles. Now came de Broglie and proposed that electrons and all other particles were in fact waves. Einstein was enthusiastic: "He has lifted a corner of the great veil." This marked the birth of the infamous "wave-particle duality" that underlies quantum uncertainty with all its consequences. Any particle, as long as not measured, resides it several locations or states. It is superposed. Einstein famously hated this term, yet he was honest enough to regard de Broglie's discovery as lifting Nature's veil.

In what follows we suggest that time is ripe for this veil to be lifted even further, strictly within present-day quantum theory. Based solely on ordinary quantum measurements, albeit under special post-selections, we show that what has been so far grudgingly accepted as "a particle existing in several locations" may in reality be a plethora of states, so far unnoticed. These states are sometimes very strange, *e.g.*, "a particle disappearing in one location and reappearing in another." It is post-selection, a simple method having a crucial role in the Two-State Vector Formalism (TSVF), which offers us a glimpse into this deeper reality beneath the quantum realm, with much more lying ahead.

This article's outline is as follows. In Secs. 3,4 we review the three-boxes paradox and its "shutter version", respectively. In Sec. 5 we present a novel version of this experiment where a particle seems to disappear and then re-appear in another place. In Sec. 6 this disappearing and re-appearing particle acts as a shutter, thereby making the paradox more tangible. Secs. 7,8 put these thought experiments in a broader context, relating them to some other ideas and Sec. 9 concludes the work.

## 3. The three-boxes paradox

The three-boxes paradox [1] is a simple and by now well-known paradox, but it already contains some intriguing features we shall rely on.
A system is prepared, i.e. pre-selected, at the superposition state:

$$|\psi(t=0)\rangle = \frac{1}{\sqrt{3}}(|1\rangle + |2\rangle + |3\rangle). \quad (1)$$

Later in time, it is (strongly) measured, i.e. post-selected, and found at the state:

$$|\phi(t=t_f)\rangle = \frac{1}{\sqrt{3}}(|1\rangle + |2\rangle - |3\rangle). \quad (2)$$

A surprising prediction arises as to measurements at intermediate times: If the particle is looked for in Box 1 at some $0 < t' < t_f$, it will be found there with certainty. To understand why this is the case, let us assume that we look for the particle at Box 1 and don't find it there, meaning that the initial state has now collapsed to: $|\psi(t=t')\rangle = \frac{1}{\sqrt{2}}(|2\rangle + |3\rangle)$. Since this state is orthogonal to $|\phi(t=t_f)\rangle$, the latter would not occur in contrast to our assumption, hence, the particle must be in the first box. However, if we alternatively look for it in Box 2, we always find it there based on the same logic. But how can it reside in both? An analysis in terms of weak values accordingly reveals that $\langle \Pi_1 \rangle_w = \langle \Pi_2 \rangle_w = +1$, where $\Pi_i = |i\rangle\langle i|$ is a projection on the i-th Box, while intriguingly the weak value at the third box is found to be anomalous $\langle \Pi_3 \rangle_w = -1$, meaning that the sum of all three weak values is still 1, as it should be. Since the weak values of projections into the first two boxes coincides with the eigenvalues of these dichotomic operators, it follows [1] that they could have been found using a strong measurement as well. This is not true for the last box which contains a particle with negative properties, a notion that will become clearer in Sec. 5.

## 4. The "shutter version" and its recent realization

The three-boxes paradox is based on a counterfactual: "Had one measured the particle in Box 1, it would be there, and similarly for 2" This is an extravagant type of nonlocality, where the very act of measurement in a certain location seems to force the particle to "collapse" there. Yet it is a retrodiction, holding for a past state which is by definition no longer accessible. This seems to rob the paradox much of its acuity.

This obstacle was overcome by Aharonov and Vaidman [2] in the general N boxes case, with a scheme analogous to Hardy's delayed measurement [3]. This time it is a strong projective measurement of a particle operating as a "shutter." It has two stages: i) The shutter particle, after the pre-selection (1), is coupled with a superposed test particle. ii) Then, after the former's post-selection (2), the test particle is subjected to measurement which reveals the shutter particle's intermediate state. The retrodiction thus turns into a standard prediction: In all cases where the shutter's post-selection succeeds, the test particle is reflected from all the N-1 boxes, demonstrating the shutter's simultaneous existence in all of them. Okamoto and Takeuchi [4] have recently tested this prediction of the TSVF for N=3 in a quantum optics setup employing a novel quantum router [5] which enables the shutter and the probe photon to indirectly interact. Their results show that the shutter did not randomly occupy one of the two boxes, but rather (within experimental accuracy limitations) reflected a photon from both.

## 5. The disappearing and reappearing particle

Whereas the three boxes experiment presents an intriguing *state* – one particle "collapsing" in whatever box inspected – our recent disappearing-reappearing particle gedankenexperiment [6] gives even more intriguing *evolution*: The particle is bound to be found in mutually-exclusive boxes at different times. And here again, the prediction is for strong (projective) measurements.

The pre-selection (in fact a preparation) is

$$|\psi(t=0)\rangle = \frac{1}{\sqrt{3}}(|1\rangle + i|2\rangle + |3\rangle), \tag{3}$$

and the post-selection

$$|\phi(t=t_f)\rangle = \frac{1}{\sqrt{3}}(-|1\rangle + i|2\rangle + |3\rangle). \tag{4}$$

We then introduce a time evolution $H = \varepsilon \sigma_x$ during $0 \leq t \leq t_f$, allowing the particle to tunnel between Boxes 1 and 2.

Surprisingly, we now have three different predictions for three earlier instants during this interval:

$$\Pi_1(t_1) = 1, \quad \Pi_3(t_1) = 1, \tag{5}$$

$$\Pi_1(t_2) = 0, \quad \Pi_2(t_2) = 0, \quad \Pi_3(t_1) = 1, \tag{6}$$

$$\Pi_2(t_3) = 1, \quad \Pi_3(t_3) = 1, \tag{7}$$

where the weak value notation is omitted since these the above values coincide with the eigenvalues of the projectors and hence could have been found in a strong measurement too [1].

It is intriguing nevertheless that $\langle \Pi_2(t_1) \rangle_w = \langle \Pi_1(t_3) \rangle_w = -1$, suggesting that the disappearance from Boxes 1 and 2 in Eq. 6 is a result of self-cancellation (see also Sec. 8). And moreover, we thus understand that the total number of particles within these boxes is understood to be zero throughout the course of the experiment, yet their sub-division is non-trivial.

It is time-symmetry that obliges this surprising prediction within orthodox quantum mechanics. Followed from past to future, this formulation gives a reasonable account: Once you have obtained the pre-selection (3), then, *if* you get either (5), (6) or (7), your probability to get the post-selection (4) goes from 11% to 33%.
But followed conversely (as if pre- and post-selection are interchanged), the result is odd: If you did *not* perform any intermediate measurement, just obtained (3) and (4), then *all three retrodictions (5)-(7) are equally correct*!

This, of course, is a very extraordinary claim, even in comparison with other quantum effects. *How can the probe particle indicate, with certainty, the shutter particle's disappearance from one location and reappearance at another?* It would therefore be only natural to consider a more moderate, even trivial alternative: The pre- and post-selections give us an *ensemble*, termed "family of histories" by Griffiths [7], comprised of three *distinct* groups, each of which has its own two properties:

i)   Particles that went to Box 1 and yield the post-selection (4) if measured at $t_1$;

ii)  Particles that went to Box 3 and yield the post-selection (4) if measured at $t_2$;

iii) Particles that went to Box 2 and yield the post-selection (4) if measured at $t_3$.

No position changes, then, because each particle has only one history: If found in Box 1/2/3, at any time, then it has been there *all along*.

Disproving this alternative is straightforward. Being "all along" somewhere is by definition a *local hidden variable*. Here the abovementioned Bell-Hardy proof [3] becomes crucial: Our single particle's wave-function, split into three, is as nonlocal as an EPR pair or GHZ triplet. Here too, then, any hidden variable must be nonlocal, this time extremely so to the point of obliging the particle to abruptly jump from one trajectory to another.

**6. The disappearing-reappearing particle made to act as a shutter**
Based on the Okamoto-Takeuchi [4] realization of Aharonov and Vaidman's protocol [2], our proposal for a realization of the disappearing-reappearing particle naturally follows [8].

The main new element added to the experiment is a temporal superposition preceding the spatial one. Split an initial photonic beam into three equal-intensity beams (see *e.g.* [9]) and delay the 2nd and 3rd beams such that the three parts pass through the 3-boxes system at $t_1$, $t_2$ and $t_3$, respectively.

At each instant, direct the 1/3 ray to one or more boxes in accordance with the TSVF prediction (Fig. 1) about the shutter-particle's position at that instant:

   i.   At $t_1$, the first 1/3 beam is split again by a simple BS into 2 and goes to Boxes 1+3.

   ii.  At $t_2$, the second 1/3 goes only to 3.

   iii. At $t_3$, the last 1/3, again split by simple BS, goes to 2+3.

Then, after passing all the boxes at all times, the photon is reunited in a precisely reverse manner to the above splits, first spatially and then temporally. Once all splits are completely undone, we measure the photon to see whether its initial quantum state is restored by interference.

Notice the required care: *Only this specific combination space and time paths* – $(1+3)(t_1)$, $3(t_2)$, $(2+3)(t_3)$ – will restore the photon's initial state.

In addition to this direct test of the shutter-particle's presence, there is a complementary test, easier but less decisive: *Send the photon through the boxes predicted to be* empty *at $t_2$*. This, conveniently, does not involve complicated momentum exchange between shutter and photon. Here too, the specific combination of (1+3)($t_2$) – restores the photon's initial state (and probabilistically also the combination 2($t_1$) and 1 ($t_3$)).

Earlier, with the *gedanken* version of the disappearing-reappearing particle, we used the Bell-Hardy proof against the "all along" local account. This involved a counterfactual: two parts of the shutter-particle's wave *could* as well interact with two atoms, which *would* then become entangled, manifesting a Bell-test violation.

The present setting offers us a stronger, *active* test of this kind. As Okamoto and Takeuchi [4] point out, the shutter and the test-particle become *EPR entangled*. Why, then, not subject them to a Bell-test? Violations of Bell's inequality are then bound to emerge, indicating that the "which path" outcome of each particle is nonlocally determined by the other particle's choice of measurement. Consequently, the "all along" alternative is ruled out in favor of the disappearance-reappearance account.

Fig. 1 shows a simplified test of this kind, where each particle's wave-function is split into two halves within an interferometer. The two interferometers cross each other on both arms. In 1/2 of the cases, each particle would turn out to have taken the path not crossed by the other, thereby detected by one of the detectors placed farther on its non-collision path in Fig. 1. Let these cases be selected out. The remaining cases, where the detectors remain silent, assure us, by Interaction-Free-Measurement [10], that *collision between the two particles has occurred, in superposition, on all four intersecting MZI paths*.

This is actually our post-selection. It gives rise to a full entanglement between the shutter and the test-particle:

$$|\psi\rangle = \frac{1}{\sqrt{2}}\left(|1\rangle|1\rangle + |2\rangle|2\rangle\right), \tag{8}$$

enabling an Alice and Bob to subject them to EPR-Bell measurements where each party randomly chooses between "which-path" and interference measurements. By Bell [11], each outcome is determined also by the type of measurement chosen for the other, distant particle. So much for the "all along" alternative.

### 7. From superposition to quantum entanglement – the quantum liar paradox

We recently suggested another paradox [12] belonging to the same family. Two atoms are prepared, one excited and one ground: $|\psi\rangle = |e\rangle|g\rangle$ We assume that a photon emitted from the first would necessarily excite the second. We wait the half-life time and now, apart from a few technicalities described in [12] we are sure that the atoms are entangled

$$|\psi\rangle = \frac{1}{\sqrt{2}}\left(|e\rangle|g\rangle + |g\rangle|e\rangle\right). \tag{9}$$

However, we can now run a Bell-test to verify the nonlocal correlation between them, but now a paradox ensues: In all cases where the first atom is still excited (i.e. oblivious Eq. 9), we believe that no photon was emitted, but in this case, how can they be entangled and exhibit nonlocal correlations? It turns out that the mere *possibility* of photons emission has the power of entangling the atoms and create nonlocal correlations between them. In [12] we continued in this line of thought to analyze the multi-particle scenarios (which are even more striking) and also discussed a verification using weak measurements.

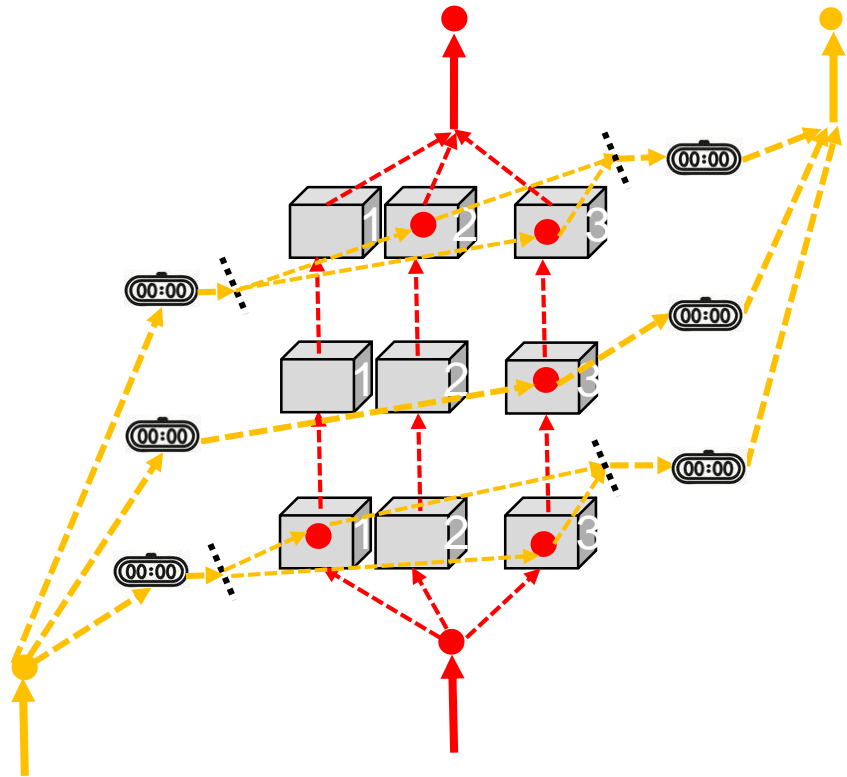

**Figure 1.** The measurement of the disappearing and re-appearing shutter is "strong" but delayed. The particle in red operates as a shutter while being superposed in Boxes 1-3. Its varying location is measured by a photon (orange) whose wave-function is split into five parts in both space and time: (1+3) at $t_1$, (3) at $t_2$ and (2+3) at $t_3$. Finally it is re-united in a reverse manner to test its interference-like pattern. This is guaranteed only with this precise combination of spacetime paths.

**8. Quantum Oblivion**

Our final task is to show that all the above effects share something in common. They all demonstrate the extreme reversibility governing quantum phenomena, to the point of events happening and then "unhappening." This is Quantum Oblivion [13], presented in greater detail DICE 2014 [14]. Moreover, it offers the very key for understanding numerous other quantum oddities.

The thought experiment in the last section demonstrates very lucidly this oblivion. The Quantum Liar paradox is based on couternfactuals, i.e. events that *could* have occurred, but eventually *haven't*. In terms of TSVF [15-17], this unique pair of happening and unhappening can be decomposed into a sum of self-cancelling *weak values* [18]. This mechanism has been shown to be ubiquitous in the quantum realm [13, 18], but here we can straightforwardly see its dynamics in action – the disappearance of a particle that should have resided in the first or second box, but eventually did not, is understood through the null weak value in these boxes which is a sum of earlier +1 and -1 weak values. The mathematical formalism leads us to recognize them as two particles, one with "positive properties" and one with "negative properties" (as can be inferred thorough weak measurements [19, 20]) which strictly cancel one another at $t_2$ thus representing a genuine disappearance, yet with an interesting underlying reality. To emphasize the latter point we suggest verification tests [6] sensitive to a non-zero modular momentum within the first and second box even in the absence of the particle.

Quantum oblivion has been shown [13] to underlie many well-known phenomena, from Interaction-Free-Measurement [10, 21] and the Quantum Zeno effect [22] to the Aharonov-Bohm effect [23]. They all turn out to share the common denominator of Nature herself totally "forgetting" a certain event, leaving an intriguing gap in causality and locality. Further work along this promising line of research are currently underway.

## 9. Discussion

We have seen a major progress in the gedanken and practical levels of quantum paradoxes. Recent pre- and post-selected paradoxes [6, 12, 21, 24] become more and more acute and can be examined using strong (projective) measurements, and not only via weak measurements. Moreover, during the last few months a laboratory demonstration [3] has achieved just that, and another one is currently under feasibility study [8]. All these broaden the notion of quantum superposition, so as to include exotic states, thus shedding light on the very foundations of quantum mechanics.

**Acknowledgments**
It is a pleasure to thank Yakir Aharonov for many helpful discussions. E.C. was supported by ERC - AdG NLST.